\documentclass[11pt,sigplan,nonacm]{acmart}
\usepackage[utf8]{inputenc}
\usepackage{natbib}
\usepackage{doi}
\usepackage{to-be-determined}
\usepackage[capitalize]{cleveref}
\usepackage{array}
\usepackage{tabularx}
\usepackage{booktabs}
\usepackage{tikz}

\title{Programmers Prefer Individually Assigned Tasks vs. Shared Responsibility}

\author{Adela Krylova}
\orcid{0000-0001-9814-9759}
\email{a.krylova@innopolis.university}
\affiliation{\institution{Innopolis University}\city{Innopolis}\country{Russia}}

\author{Roman Makarov}
\orcid{0009-0009-3364-0497}
\email{o.makarov@innopolis.university}
\affiliation{\institution{Innopolis University}\city{Innopolis}\country{Russia}}

\author{Sergei Pasynkov}
\orcid{0000-0001-0000-0000}
\email{s.pasynkov@innopolis.university}
\affiliation{\institution{Innopolis University}\city{Innopolis}\country{Russia}}

\author{Yegor Bugayenko}
\orcid{0000-0001-6370-0678}
\email{yegor256@gmail.com}
\affiliation{Huawei\city{Moscow}\country{Russia}}


\begin{abstract}
In traditional management, tasks are typically assigned to individuals, with each worker taking full responsibility for the success or failure of a task. In contrast, modern Agile, Lean, and eXtreme Programming practices advocate for shared responsibility, where an entire group is accountable for the outcome of a project or task. Despite numerous studies in other domains, the preferences of programmers have not been thoroughly analyzed. To address this gap, we conducted a survey featuring seven situational questions and collected the opinions of 120 software development practitioners. Our findings reveal that programmers prefer tasks to be assigned to them on an individual basis and appreciate taking personal responsibility for failures, as well as receiving individual rewards for successes. Understanding these preferences is crucial for project managers aiming to optimize team dynamics and ensure the successful completion of software projects.
\end{abstract}

\begin{document}

\maketitle

\section{Introduction}

Traditional management theory posits that workers' commitment is higher if tasks are assigned to them individually, with each individual taking full responsibility for failure and success alike~\citep{drucker2006executive, drucker2006practice,welch2011,iacocca1984}. Contrary to this ``archaic'' philosophy, proponents of lean management practices argue that assigning responsibility for failures and successes to the entire team is a more effective way of managing programmers~\cite{beck2000extreme, siakas2007agile, yu2014understanding}. Despite the discovery of ``social loafing'' in 1913~\citep{ingham1974ringelmann, kravitz1986ringelmann}, the debate remains unresolved~\citep{mccarthy2019towards}: Who should bear responsibility for a missed deadline, an individual programmer or the entire team? In other words, is individual or group responsibility, defined as ``an obligation to satisfactorily perform or complete a task''~\citep{mcgrath2018accountability}, more conducive to better management of software teams?

This debate is not purely metaphysical or philosophical~\citep{beecham2006protocol, beecham2008motivation}. Its outcome has practical implications on management practices in software development teams. If we prove that individual responsibility is more preferred by programmers, we may, for example, consider restructuring compensation mechanisms of programmers in order to reward them for smaller individual contributions (on per-task basis), rather than for larger results of the entire team (annual bonuses)~\citep{patterson2020bringing,bugayenko2020blog0623}. We may also reconsider the ``only positive encouragement'' paradigm and enable certain forms of punishment for personal failures, like missed deadlines.

The research questions that we ask in this study are the following:
\begin{description}
    \item[RQ1] To what extent do software developers perceive themselves as solely responsible for the success or failure of the tasks assigned to them?
    \item[RQ2] How does the level of professional development influence a software developer's understanding and acceptance of responsibility?
\end{description}

In order to answer these questions we created a survey of seven situational questions. Each question imaginatively places a respondent into a situation where he/she must either take a responsibility for his/her actions or shift it to the group. We tried to structure the questions such a way that a respondent does not understand the intent of our research. We distributed the questionnaire among 120 programmers, spanning various experience levels.

The survey results identified three key factors influencing programmers' responsibility within a team setting: individual task ownership, clarity of task definitions, and the role of external factors. Individual ownership fosters accountability, while clear task descriptions enhance engagement. Moreover, we discovered that professional experience does not significantly impact responsibility perception.

The rest of the paper is organized as following.
\Cref{sec:related} shows a few previous studies related to the problem we research.
\Cref{sec:method} explains our method of study.
\Cref{sec:results} presents the results we obtained.
\Cref{sec:discussion} discusses our findings and \cref{sec:limitations} highlights limitations of our method.

\section{Related Work}
\label{sec:related}

In software development, understanding how responsibility works among developers is crucial for successful project management \citep{demarco2013peopleware, franca2008qualitative, brooks1975mythical}. While numerous studies have explored different aspects of collective vs. individual responsibility~\citep{narveson2002collective,doorewaard2002team}, there exists a notable research gap in the investigation of critical factors influencing responsibility within software developers~\citep{celepkolu2018importance}. This chapter reviews existing literature to highlight the scarcity of research addressing responsibility within the developer community, providing a foundation for our investigation.

\citet{Deak_2016} conducted a study where they described motivating and demotivating factors for testers. Factors that can have a detrimental impact on motivation include time constraints, lack of recognition, ineffective management, and strained relationships. On the other hand, both extrinsic and intrinsic elements contribute to positive motivation.

\citet{Dorairaj_2012} provided an analysis of the factors leading to a lack of trust in distributed Agile teams, thus negatively impacting responsibility. These factors include the absence of a sense of belonging, feelings of vulnerability, inadequate team bonding, insufficient cultural understanding, the absence of face-to-face interaction, and ineffective communication.

Study by \citet{Vishnubhotla_2019} explores capability measurement in Agile team formation. Survey results from 60 practitioners align with the SLR, identifying 127 individual and 28 team capability measures. Notably, seven individual and one team measure were newly identified.

\citet{Austin_2001} presented an agency model that investigates the effects of time pressure on software quality in the context of software development. The findings suggest that systematically adding slack to deadlines may not always be an effective strategy for minimizing costs, and that setting deadlines separately from planning estimates can lead to improved outcomes.

To our knowledge, there has not been any systematic study of an individual vs. team responsibility dichotomy. It still remains unclear whether programmers prefer to be personally responsible for their tasks or work with jobs assigned to their teams.

\section{Method}
\label{sec:method}

We created a survey method to gather responses about the attitudes toward responsibility among a community of software developers, spanning various experience levels. We designed a survey utilizing Google Forms, an online platform designed for executing surveys.

The first question in our survey served for confirming that the respondents belongs to the target group. Following seven situational questions (\cref{tab:tab1}) were directly related to our research questions. Then, final three questions collected demographic profile of respondents. We ended the survey with a non-mandatory open-ended question. All questions were written in English.

The seven situational questions (\cref{tab:tab1}) were intentionally designed to conceal the purpose of the survey. We expected respondents to imagine work situations and to respond to them honestly, thus making it possible for us to understand what are their most probable behavior would be.

\newcommand{\results}[4]{
    \renewcommand{\arraystretch}{1}
    \begin{tabular}[t]{@{}r@{~}r@{}}
    \score{1} & #1\% \\
    \score{2} & #2\% \\
    \score{3} & #3\% \\
    \score{4} & #4\% \\
\end{tabular}}
\newcounter{question}
\renewcommand\thequestion{{\sffamily\bfseries Q\arabic{question}}}
\makeatletter
\newcommand{\question}[2][]{{\sffamily\bfseries Q{}\refstepcounter{question}\label@noarg{Q:#1}\arabic{question}}: ``#2''}
\makeatother
\newcommand{\score}[1]{{\ttfamily[#1]}}
\begin{table*}\small\renewcommand{\arraystretch}{1.3}
\begin{tabularx}{\textwidth}{
  >{\raggedright}X % question
  >{\raggedright}p{0.075\linewidth} % type
  >{\raggedright}p{0.2\linewidth} % answers
  >{\raggedright}p{0.2\linewidth} % interpretation
  >{\raggedright\arraybackslash}p{0.1\linewidth}} %results
\toprule
Question & Type & Possible answers & Interpretation & Results\\
\midrule
\question[assign]{You have two assignments to complete during this working day: one that was assigned to you personally, another that you work on with your team. Which one would you start with?}
    & \score{1}--\score{4} Likert Scale
    & \score{1} for the first task and \score{4} for the second task
    & Answers closer to \score{4} contribute toward the strong sense of individual responsibility (RQ1)
    & \results{40.3}{19.3}{25.2}{15.1}\\
\question[success]{A project that you worked on during your probationary period has finished successfully. How likely do you think that you will be offered to join another project?}
    & \score{1}--\score{4} Likert Scale
    & \score{4} means very likely and \score{1} means very unlikely
    & Responses close to \score{1} show smaller responsibility for the success of the project, probably due to lower experience level (RQ2)
    & \results{1.7}{12.6}{38.7}{47.1}\\
\question[failure]{A recent project, which you were a part of, was closed by the company due to unmet deadlines and incomplete functionality. What should be done?}
    & Closed question
    & \score{1} if project managers should be fired;
    \score{2} if development team;
    \score{3} if the whole team;
    \score{4} if no one should be penalized
    & \score{4} means preference of group responsibility (RQ1 + RQ2)
    & \results{56.1}{0.0}{18.2}{25.8}\\
\question[laptop]{Suppose your laptop broke down and you were not able to finish your work on time. Do you think that it is right to penalize you in this case?}
    & \score{1}--\score{4} Likert Scale
    & \score{1} for ``absolutely wrong'' and \score{4} for ``absolutely right''
    & Answers close to \score{4} show strong individual responsibility (RQ1)
    & \results{36.4}{29.5}{19.7}{14.4}\\
\question[client]{Given two tasks, you can't complete both due to a personal fault. The first task is crucial for the client, but you won't suffer much personally if it's delayed. The second task carries a penalty for missing the deadline, though it's less important for the overall project. Which task would you prioritize?}
    & \score{1}--\score{4} Likert Scale
    & \score{1} for the first task and \score{4} for the second task
    & Answers close to \score{1} show stronger responsibility towards client satisfaction (RQ1), while \score{4} shows low personal responsibility
    & \results{37.9}{18.2}{12.9}{31.1}\\
\question[unclear]{You got a task with an unclear description. Because of your mistake, you noticed it just before the deadline. Would you finish it the way you think is right without asking for help?}
    & \score{1}--\score{4} Likert Scale
    & \score{1} means doing the task on your own and \score{4} means asking for clarification
    & Answers close to \score{1} show lower personal responsibility (RQ1)
    & \results{11.4}{13.6}{20.5}{54.5}\\
\question{You have a project part to complete, but you can only start when your colleague finishes theirs. They say they'll be late by two days and ask you not to inform the manager. What would you do?}
    & \score{1}--\score{4} Likert Scale
    & \score{1} means informing project manager and \score{4} means letting programmer do their job
    & \score{1} shows that the individual responsibility is more important for the respondent, while \score{4} means waiting, hence loosing resources (RQ1)
    & \results{42.4}{26.5}{16.7}{14.4}\\
\bottomrule
\end{tabularx}
\caption{All questions from our online survey, together with possible answers for each question, and the actual results collected (in the last column).}
\label{tab:tab1}
\end{table*}

\section{Results}
\label{sec:results}

This section presents the results obtained from our survey, while the entire collected dataset is available in open GitHub repository\footnote{\url{https://github.com/rmakarovv/Responsibility_Study}}. Upon gathering responses, we analyzed participants' work experience, scrutinizing the distribution patterns after classifying participants into two groups: the first group comprising junior and middle programmers, and the second consisting of senior software developers. Our categorization was done that based on the assumption that there could be a gap in working habits between beginners and advanced specialists.

Consciously opting for a modest number of questions, we aimed to make the survey interesting and easy to navigate for the respondents, ensuring a straightforward completion process. This decision proved to be successful, evidenced by a high completion rate—every respondent who initiated the survey concluded the entire questionnaire.

We recruited respondents from posts on social networks, and in total, we gathered responses from 120 individuals. All participants completed all mandatory questions, with 20 of them providing answers to the final open-ended question. Upon analyzing the work experience of our respondents, we discovered that all our respondents have a solid experience as a programmers except one. Only one respondent did not prove to belong to our target group (``programmers'') Hence, we decided to exclude this particular response.

Our respondents disclosed the following information about themselves:
\newcommand{\quest}[9]{
\item ``#1''
\newline
#3 (#2\%);
#5 (#4\%);
#7 (#6\%);
#9 (#8\%).}

\begin{itemize}
\quest{How much code did you write during the last three months?}
    {70.6}{More than 1000 lines}
    {23.5}{Up to 1000 lines}
    {5.9}{Up to 100 lines}
    {0.0}{Nothing}
\quest{What is your professional level?}
    {52.1}{Senior}
    {31.9}{Middle}
    {5.0}{In my field there is not such a segmentation}
    {10.9}{Junior or internship}
\quest{Do you currently work?}
    {86.6}{Yes, full-time}
    {3.4}{No, but I have some working experience}
    {4.2}{Currently searching for a job}
    {5.9}{Yes, part-time}
\end{itemize}

Our research questions were framed as ``To what extent do software developers perceive themselves as solely responsible for the success or failure of the tasks assigned to them?'' and ``How does the level of professional development influence a software developer's understanding and acceptance of responsibility?.'' Based on these questions, we designed a survey that is presented in \cref{tab:tab1}, along with the response distribution.

\subsection{RQ1: To what extent do software developers perceive
themselves as solely responsible for the success or failure of
the tasks assigned to them?}

The results of \ref{Q:assign} show that 69.6\% of workers feel more responsible for completing tasks assigned solely to them. Similarly, \ref{Q:success} reveals that a majority of developers (85.8\%) tend to take responsibility for the success of a project. These findings suggest that individual ownership is crucial for fostering a sense of accountability. However, in teamwork environments where tasks are shared, developers may feel that their work could be done by someone else, leading to a decreased sense of ownership and ultimately, a lower perception of personal responsibility.

The results of \ref{Q:unclear} revealed that clearly defined tasks can significantly enhance developers' sense of responsibility.

\subsection{RQ2: How does the level of professional development
influence a software developer’s understanding and acceptance of responsibility}

After collecting responses, we divided them into two categories based on participant experience: junior/middle developers and senior developers. Analyzing their responses revealed similar distributions for both groups across all questions. Thus, contrary to our initial expectations, the findings did not reveal a significant difference in the perception of responsibility between less experienced and more experienced developers. This suggests that the underlying trait of responsibility is primarily personal and not significantly impacted by age or professional experience. While professional development can enhance knowledge and skills, it may not significantly impact an individual's innate sense of responsibility. Hence, answering the RQ2, the level of professional development does not influence a software developer’s understanding acceptance of responsibility.

\section{Discussion}
\label{sec:discussion}

The focus of this study was to explore the perception of responsibility among software developers, with a particular emphasis on the distinctions among developers of different experience levels. The findings reveal several key factors influencing the sense of responsibility in these two groups.

\textbf{Why the professional experience of a programmer does not impact the responsibility for an assigned task?}
We believe that the correlation between professional experience and responsibility for assigned tasks is not absolute due to several factors. Firstly, experience alone does not guarantee the acquisition of essential soft skills, such as effective communication, adaptability, and a strong work ethic, which are crucial for assuming responsibility. Individuals may accumulate years of technical experience but lack the interpersonal and organizational skills necessary for responsible task handling. Secondly, work environments and cultures vary, and some organizations may foster a sense of responsibility at all experience levels, emphasizing individual qualities over tenure. Additionally, the dynamic nature of the technology field requires continuous learning and adaptation, and a less experienced but proactive individual may demonstrate greater responsibility by actively seeking knowledge and staying current.

\textbf{How do software developers perceive their responsibility in the face of external factors that hinder their work?}
Most respondents feel responsible for the results of their work. However, if some outer factors hinder their work, such as broken laptop (\ref{Q:laptop}), they do not feel responsible for missing the deadline. While developers generally accept responsibility for their work (\ref{Q:client}), they may attribute a portion of the responsibility for missed deadlines to external factors like equipment malfunctions, as we found out that 47.9\% of respondents would not feel responsible for failing the task if their laptop broke down (\ref{Q:laptop}). Hence, organizations should acknowledge this perspective and provide support mechanisms to help developers mitigate the impact of these challenges.

\textbf{Why does Agile emphasize team responsibility, while the opinion of developers show that people prefer individual responsibility?}
Contemporary software development reached high level of complexity; tasks often require collaboration of several specialists---that makes it impractical to assign them to individuals. We believe that Agile underscores shared responsibility as a response to the need for collaborative efforts in addressing multifaceted challenges. In the context of software development, where a diverse skill set is crucial, a collective, team-oriented approach becomes essential. Nevertheless, it is crucial to recognize the individual nature of preferences, as there are scenarios where assigning individual responsibility may indeed prove more effective, depending on the task's nature and an individual's specific strengths.

\textbf{How Can Management Help Programmers Become More Responsible for Their Tasks?}
First, as programmers prefer working on tasks assigned to them individually, according to \ref{Q:assign}, management should strive to decompose the project scope into smaller elements and then assign tasks to programmers on an individual basis.
Second, because programmers appreciate it when both success (\ref{Q:success}) and failure (\ref{Q:failure}) are attributed personally rather than at a group level, management should establish motivation mechanisms where both positive and negative incentives are distributed based on individual achievements~\cite{delfgaauw2020team}.
Third, as programmers do not appreciate being punished for failures, according to \ref{Q:laptop}, management should lean more towards mechanisms of positive encouragement.
Fourth, given that programmers, according to \ref{Q:unclear}, do not hesitate to seek clarification on the scope of tasks assigned to them, management should facilitate open communication between programmers and product owners, thus enabling earlier refinement of requirements, especially in distributed teams.
There could be additional recommendations for managers along these lines.

\section{Limitations}
\label{sec:limitations}

It is essential to recognize the limitations of this study. With a sample size of 120 respondents, the study's scope is relatively narrow. Additionally, the respondents were primarily drawn from personal connections, with a significant portion being Russian individuals. This sampling approach could limit the generalizability of the findings to the wider population of software developers. Expanding the sample size and diversifying the respondent pool could provide a more thorough understanding of responsibility perceptions across the broader software development community.

\section{Conclusion}
\label{sec:conclusion}

This study looked closely at how software developers see responsibility, focusing on differences between those with more and less experience. We found that programmers feel more responsible for tasks assigned to them individually rather than to the group to which programmers belong. Surprisingly, the level of experience did not show a big difference in responsibility perception, suggesting it is a personal trait unrelated to professional growth. Recognizing study limits, such as a small sample of participants from different backgrounds, cautions against making broad conclusions. However, we depicted significant factors and answered set research questions. Moreover, we provide a foundation for future research in the field of software development.

\section{Acknowledgement}

We would like to thank Nikita Bogdankov, Dilyara Farkhutdinova and Alexey Potyomkin for their contributive review of our draft.

\bibliographystyle{ACM-Reference-Format}
\bibliography{refs}

\end{document}